\documentclass[eqsecnum,showpacs,aps]{revtex4}
\usepackage{epsfig}

\newcommand{\beq}{\begin{equation}}
\newcommand{\eeq}{\end{equation}}
\newcommand{\beqa}{\begin{eqnarray}}
\newcommand{\eeqa}{\end{eqnarray}}
\def\ket#1{|\,#1\,\rangle}
\def\bra#1{\langle\, #1\,|}

\def\proj#1#2{\ket{#1}\bra{#2}}
\def\expect#1{\langle\, #1\, \rangle}

\def\ol#1{\overline{#1}}
\def\kpsi{\ket{\psi}}

\def\half{\frac{1}{2}}
\def\opone{\leavevmode\hbox{\small1\kern-3.8pt\normalsize1}}

\begin{document}
\title{Broadcasting of entanglement at a distance using
linear optics and telecloning of entanglement }

\author{Iulia Ghiu$^{1,2}$}
\email{iughiu@barutu.fizica.unibuc.ro}
\author{Anders Karlsson$^2$}
\email{andkar@imit.kth.se}
\affiliation{$^1$Department of Physics, University of Bucharest,
P.O. Box MG-11, R-077125, Bucharest-M\u{a}gurele, Romania}
\affiliation{$^2$Department of Microelectronics and Information Technology,
Royal Institute of Technology (KTH),
 Electrum 229, 164 40 Kista, Sweden}

\date{\today}

\begin{abstract}
We propose a scheme for broadcasting of
entanglement at a distance based on linear optics.
We show that an initial polarization entangled state can be simultaneously split and transmitted to a pair of observers situated at different locations with the help of two conditional Bell-state analyzers based on two beam
splitters characterized by the same reflectivity $R$. In particular for $R=1/3$
the final states coincide with the output states obtained by the
broadcasting protocol proposed by Buzek {\it et al.} [Phys.
Rev. A {\bf 55}, 3327 (1997)]. Further we present a different protocol
called telecloning of entanglement, which combines the
many-to-many teleportation and nonlocal optimal asymmetric cloning
of an arbitrary entangled state. This scheme allows the optimal transmission of the two nonlocal optimal clones of an entangled state to two pairs of spatially separated receivers.
 \end{abstract}

\maketitle

\section{Introduction}
The concept of entanglement plays a central role in many processes
of quantum information theory: quantum teleportation
\cite{Bennett}, quantum telecloning \cite{Murao1},
telebroadcasting of entanglement \cite{Ghiu}, quantum cryptography
\cite{BB,Ekert,Hillary}, entanglement swapping
\cite{Zukowski,Bose}, quantum computation \cite{Shor}. With the
basic teleportation protocol useful as such in quantum
communications and quantum computing, some generalizations of
quantum teleportation have been considered: one-to-many
\cite{Murao2} and many-to-many \cite{Ghiu}, where the information of a quantum system
is distributed from one sender to many receivers, and from $N$
senders to $M$ receivers $(M>N)$, respectively. The experimental
implementations of quantum teleportation \cite{Bouw} and
entanglement swapping \cite{Pan,carte} were realized by performing
conditional Bell-state measurements based on balanced beam
splitters.

One main difference between the classical and quantum information
theory is the no-cloning theorem, which imposes that an arbitrary
quantum pure state cannot be copied \cite{Wootters}. There is an
extension of this theorem for mixed states \cite{Barnum}, where
Barnum {\it et al.} have shown that one cannot broadcast two
noncommuting mixed states. Because the perfect copying of a pure
state is impossible, cloning machines were considered for
generating two or more identical mixed output states
\cite{Buzek,Bruss,Buzek3,Gisin,Bruss2,Werner,Keyl,Zanardi,Fan}.
Cerf has introduced the $1\to 2$ asymmetric cloner that produces
two states emerging from two different Heisenberg channels
\cite{Cerf1,Cerf2}. Lamoureux {\it et al.} have proved
that no quantum operation exists that copies the entanglement of
all maximally qubit pairs (``entanglement no-cloning principle'')
\cite{Lam}. Recently, a partial teleportation scheme was
introduced by Filip \cite{Filip1,Filip2}, which is performed using an unbalanced beam splitter with the reflectivity $R$. This protocol can be also
viewed as an optimal universal asymmetric cloning at a distance.
Experimentally, both asymmetric cloning and telecloning of single photons were demonstrated using type-II downconversion by Zhao {\it et al.} \cite{Zhao}.
An interesting aspect of quantum cloning is that it can be used for broadcasting of entanglement into two identical inseparable states
\cite{Buzek2,Buzek3}. One of us has analyzed the
asymmetric broadcasting of inseparability using both local and
nonlocal optimal cloning machines \cite{Ghiu}. In a recent paper Demkowicz-Dobrzanski {\it et al.}  have proved that classical communication plays a crucial role in the process of broadcasting of entangled states, because it improves the fidelity of the output states for a certain class of initial entangled states \cite{Rafal}.
We want to emphasize that the process of broadcasting of entanglement requires all the particles (the original ones, the copies, and the ancillas) to be shared only by two distant observers.

In this paper we propose two schemes for splitting of entanglement at a distance: the first one is an extension of the partial teleportation of qubits, while the second one is a generalization of the results obtained in Ref. \cite{Ghiu},
as well as closely related to the so called cloning of quantum
registers \cite{Buzek3}. First we answer the following question: Is there a
local protocol that allows Alice and Bob to broadcast the
entanglement in such a way that they keep one final pair and send
the other one to two spatially separated observers, Charlie and
Daniel, simultaneously? We show that this is possible by proposing
a scheme, which requires two conditional Bell-state analyzers.
Suppose that Alice and Charlie, and Bob and Daniel, share two
singlet polarization states. Alice and Bob mix locally their
photons on two unbalanced beam splitters spatially separated. We
prove that for some specific values of the reflectivity $R$ of the
beam splitters, the inseparability is asymmetrically broadcast at
a distance. We find that for $R=1/3$ the two final states are
identical and also coincide with the two output states given by
the local broadcasting protocol introduced by Bu$\check z$ek {\it
et al.} \cite{Buzek2}. Our scheme may be viewed as a double conditional
entanglement swapping protocol.

In the last few years much progress has been made in the study of telecloning of a state of a quantum system situated in one location: symmetric telecloning of qubits \cite{Murao1,Dur}, and qudits \cite{Murao2}, and asymmetric telecloning of qubits \cite{Murao2}, and qudits \cite{Ghiu}. The second scheme proposed in this paper presents the telecloning of an entangled state shared by two parties situated at two different locations.
This protocol performs
simultaneously the many-to-many teleportation and nonlocal optimal
asymmetric cloning of inseparability. Suppose that two spatially separated persons $A_1$
and $A_2$ share an unknown
entangled state and they wish to send two copies to two pairs of
observers $B_1$-$B_2$ and $B_3$-$B_4$. The channel required in our protocol is an
eight-particle maximally entangled state. Each sender performs a
Bell measurement and communicates the outcomes to the receivers,
who apply local unitary operators, generating the optimal
asymmetric clones.
We have proposed in Ref. \cite{Ghiu} a scheme called
telebroadcasting, a combination of many-to-many teleportation and
asymmetric broadcasting of entanglement. The final states in the
two processes, telecloning and telebroadcasting of entanglement,
are distributed between two pairs of observers, but they are
different, namely they are obtained after tracing over the ancillas on the states generated by applying nonlocal optimal universal
asymmetric cloning machines, and local optimal universal
asymmetric cloning machines, respectively.

The paper is structured as follows.
In Sec. II A we review the scheme proposed by Filip for the optimal asymmetric cloning at a distance of a qubit, and then we find the relation between the reflectivity of the beam splitter used in this protocol and the parameter ``$p$'', which characterizes a universal optimal asymmetric cloning machine. Further, in Sec. II B, we propose a scheme for broadcasting of entanglement at a distance based on two conditional Bell-state analyzers consisting in two beam splitters having the same reflectivity $R$. The entanglement is symmetrically broadcast for $R=1/3$, when the emerging states are the same as in the Buzek's scheme \cite{Buzek2}, where two local universal cloning machines were employed. In Sec. II C we show that, although the two final states obtained in the broadcasting of entanglement at the distance do not violate the Clauser-Horne-Shimony-Holt (CHSH)-Bell inequality, they can be used as quantum channels in the standard teleportation.
In Sec. III we present the asymmetric telecloning of entanglement,
a process which combines the many-to-many teleportation and optimal universal asymmetric cloning. Finally, our conclusions are summarized in Sec. IV. In the Appendix we analyze the inseparability of the final states obtained in the broadcasting of entanglement at a distance, which is required in Sec. II B.

%.............................................................
\section{Asymmetric broadcasting of entanglement at a distance}
\subsection{Preliminaries}
Before considering the broadcasting of entanglement, let us review the scheme proposed by Filip for the partial optimal teleportation.
In the standard teleportation protocol an unknown qubit is
transmitted from Alice to Bob \cite{Bennett}, while Alice's state
is destroyed. On the other hand, Filip has found that it is possible
to send one imperfect copy to Bob, and at the same time Alice to
get another imperfect copy of the initial state
\cite{Filip1,Filip2}. This protocol, called partial optimal
teleportation, is based on a conditional Bell-state analyzer
characterized by an unbalanced beam splitter. Filip has proved
that the fidelities between the final states and the initial one are
 \beqa
\label{fid1}\label{fFilip}
F_A&=&\frac{2-6R+5R^2}{2(1-3R+3R^2)},\nonumber\\
F_B&=&\frac{1-2R+2R^2}{2(1-3R+3R^2)},
\eeqa
 where $R$ is the
reflectivity of the beam splitter $(R\le 1/2)$. Note that the original
teleportation protocol as implemented experimentally using
beam splitters ($F_A=1/2, F_B=1$) is retrieved for $R=1/2$, while for
$R=1/3$ one obtains the universal quantum cloning $F_A=F_B=5/6$.

It is worth emphasizing that the two fidelities of Eq. (\ref{fFilip}) saturate the cloning inequality,
i.e., the fidelity of one state is maximal for a fixed fidelity of the other one \cite{Filip1,Filip2}.
For this reason, the partial optimal teleportation can be viewed also as the optimal asymmetric cloning.
We may ask now: Which is the asymmetric cloning machine that generates the same states as Filip's scheme?

One of us has obtained in Ref. \cite{Ghiu} the expression of the optimal universal asymmetric Heisenberg cloning machine,
\beqa \label{general}
&&U\ket{j}\ket{00}
=\frac{1}{\sqrt{1+(d-1)(p^2+q^2)}}( \ket{j}\ket{j}\ket{j}\nonumber\\
&&+p\sum_{r=1}^{d-1}\ket{j}\ket{\ol{j+r}}\ket{\ol{j+r}}
+q\sum_{r=1}^{d-1}\ket{\ol{j+r}}\ket{j}\ket{\ol{j+r}}),
\eeqa
where $p+q=1$ and $d$ is the dimension of the system that is cloned. For $d=2$ we get
\beqa \label{Pauli}
U(p)\ket{0}\ket{00}=\frac{1}{\sqrt{1+p^2+q^2}}(\ket{000}+p\ket{011}+q\ket{101}),\nonumber\\
U(p)\ket{1}\ket{00}=\frac{1}{\sqrt{1+p^2+q^2}}(\ket{111}+p\ket{100}+q\ket{010}).
\eeqa
The fidelities of the two output states are \cite{Ghiu}
\beqa \label{fid2}
F_A&=&\frac{2-2p+p^2}{2(1-p+p^2)},\nonumber\\
F_B&=&\frac{1+p^2}{2(1-p+p^2)}.
\eeqa
From Eqs. ({\ref{fid1}) and ({\ref{fid2}) we easily find that
\beq \label{pR}
p=\frac{R}{1-R},
\eeq
therefore the two final states obtained in the
partial optimal teleportation are generated by the cloning machine (\ref{Pauli}) $U(R/(1-R))$.

%.............................................................
\subsection{Broadcasting of entanglement at a distance}

In this section we prove that by mixing each particle of an
initial entangled state with one half of the singlet state on two
beam splitters, we can broadcast entanglement at a distance. We
obtain a family of two less entangled states which depends on the
reflectivity $R$ of the beam splitters. The two output states are
identical for $R=1/3$ and represent the two output states
generated in the broadcasting protocol introduced by Buzek {\it et
al.} in Ref. \cite{Buzek2}.

The scheme for broadcasting of entanglement at a distance is shown
in Fig. 1. Suppose two observers, Alice and Bob, share a pure
polarization entangled state of two photons, 
\beq \label{ini}
 \kpsi_{a_1b_1}=\alpha \ket{HH}+\beta \ket{VV},
\eeq 
where  $\alpha$ and $\beta $ are unknown coefficients such that $|\alpha |^2+|\beta |^2=1$. In this
section we will use the notation $\ket{0}=\ket{H}$,
$\ket{1}=\ket{V}$. They want to split it, sending one state to two
observers, Charlie and Daniel, situated at different locations,
 and at the same time to keep the second state.
Alice and Charlie share the first singlet state, and Bob and
Daniel the second singlet state. The state of the whole system is
\beqa \label{init}
&&\ket{\xi }_{in}=\kpsi_{a_1 b_1}\ket{\Psi^-}_{{a_2}c}\ket{\Psi^-}_{{b_2}d}\nonumber\\
&&=\alpha \ket{0}_{a_1}\frac{1}{\sqrt 2}(\ket{01}-\ket{10})_{{a_2}c}\ket{0}_{b_1}\frac{1}{\sqrt 2}(\ket{01}-\ket{10})_{{b_2}d}\nonumber\\
&&+\beta \ket{1}_{a_1}\frac{1}{\sqrt 2}(\ket{10}-\ket{01})_{{a_2}c}\ket{1}_{b_1}\frac{1}{\sqrt 2}(\ket{10}-\ket{01})_{{b_2}d}.
\eeqa

\begin{figure}
\includegraphics{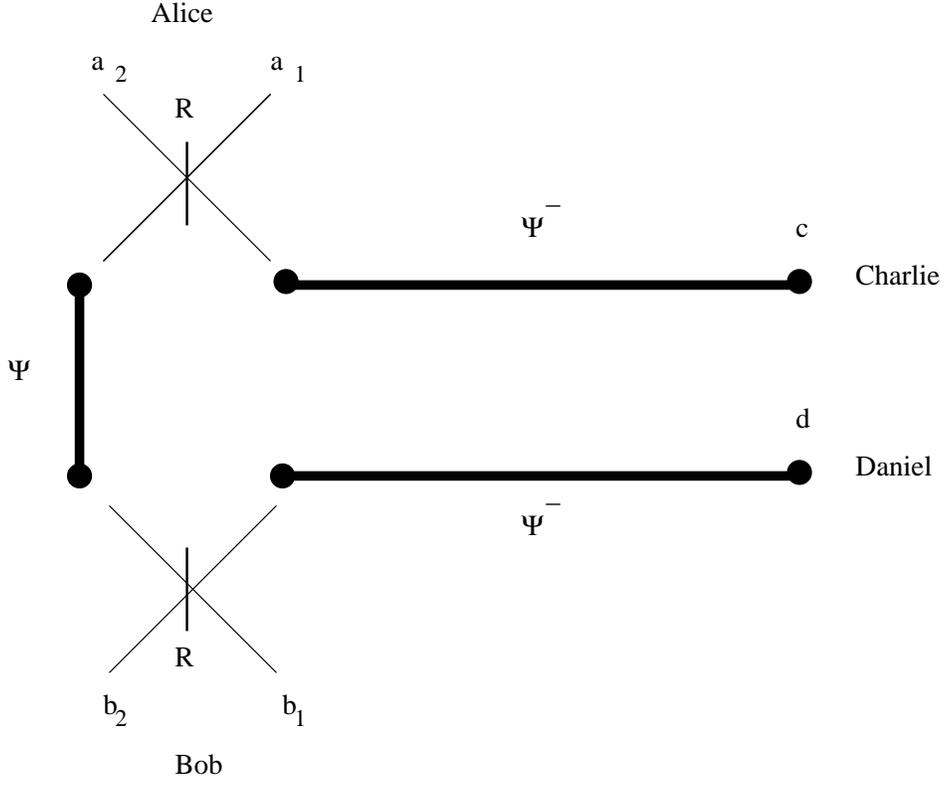}
\vspace{0.5cm}
\caption{The scheme for conditional broadcasting of entanglement at a distance. Alice and Bob share an entangled state $\kpsi $, which they wish to broadcast. In addition, Alice and Charlie, and Bob and Daniel have two singlet states distributed between them. The thick lines represent the entanglement. }
\end{figure}

Each of the two senders, Alice and Bob, mixes the photons available on a beam splitter of reflectivity $R$, where $R\le 1/2$.
The Alice's beam splitter is described by the transformation:
\beq
U^a_{BS}\ket{\psi}_{a_1}\ket{\chi}_{a_2}=\left( \sqrt T\ket{\psi}_{a_1}+\sqrt R\ket{\psi}_{a_2}\right) \left( \sqrt T\ket{\chi}_{a_2}-\sqrt R\ket{\chi}_{a_1}\right), 
\eeq
with $T+R=1$,
and we have a similar expression for Bob's beam splitter. The action of the two beam splitters on the initial state is given by
\beq \label{actionBS}
U^a_{BS}\otimes U^b_{BS}\ket{\xi}_{in}=\sqrt{\lambda_s}\ket{\phi_s}+\sqrt{\lambda_d}\ket{\phi_d},
\eeq
where $\ket{\phi_s}$ is the normalized state, which characterizes the case when the photons leave the beam splitters on the same arms. $\ket{\phi_d}$ is the normalized state obtained in the case when the photons leave the beam splitters on different arms, and this is equivalent to applying the operator $\Pi_{a_1a_2}\otimes I_c\otimes \Pi_{b_1b_2}\otimes I_d$ on the initial state, where \cite{Filip1,Filip2}
\beq
\Pi =(1-2R)I\otimes I+2R\proj{\psi^-}{\psi^-}.
\eeq

In the following we will analyze the case when the four photons leave the beam splitters separately, because only in this case the entanglement is broadcast.

If $R=\half $, and Alice and Bob share a maximally entangled state
($\alpha =1/\sqrt 2$), then our protocol becomes a conditional
entanglement swapping scheme with two swaps of entanglement at
each end of the maximally entangled initial state. In this case
Charlie and Daniel obtain a maximally entangled state,
 while the entanglement between Alice and Bob is destroyed.

The final state obtained in the interesting case when the photons leave the two beam splitters on different arms, 
 \beq
 \ket{\phi_d}=\Pi_{a_1a_2}\otimes I_c\otimes \Pi_{b_1b_2}\otimes I_d\ket{\xi}_{in},
\eeq is given by Eq. (\ref{fi}) in the Appendix.
 The initial state is broadcast if $\rho_{{a_1}{b_1}}$ and $\rho_{cd}$ are inseparable,
 while $\rho_{{a_1}c}$ and $\rho_{{b_1}d}$ are separable. In the Appendix we
 investigate under what conditions this is possible. The final states are \beqa \label{stfinala1}
&&\rho_{{a_1}{b_1}}(R)=\frac{1}{4(1-3R+3R^2)^2}[4(1-2R)^2(1-R)^2\proj{\psi}{\psi}\nonumber\\
&&+R^4\proj{00}{00}+4(1-2R)(1-R)R^2|\alpha |^2\proj{00}{00}\nonumber\\
&&+R^4\proj{11}{11}+4(1-2R)(1-R)R^2|\beta |^2\proj{11}{11}\nonumber\\
&&+R^2(2-6R+5R^2)(\proj{01}{01}+\proj{10}{10})],
\eeqa
\beqa \label{stfinala2}
&&\rho_{cd}(R)=\frac{1}{4(1-3R+3R^2)^2}[4R^2(1-R)^2\proj{\psi}{\psi}\nonumber\\
&&+(1-2R)^4\proj{00}{00}+4R(1-R)(1-2R)^2|\alpha |^2\proj{00}{00}\nonumber\\
&&+(1-2R)^4\proj{11}{11}+4R(1-R)(1-2R)^2|\beta |^2\proj{11}{11}\nonumber\\
&&+(1-2R)^2(1-2R+2R^2)(\proj{01}{01}+\proj{10}{10})].
\eeqa

Thus  we obtain that $\rho_{{a_1}{b_1}}$, $\rho_{cd}$ are inseparable,
 and $\rho_{{a_1}c}$, $\rho_{{b_1}d}$ are separable for $R$ and $\alpha $ satisfying Eqs. (\ref{condi}) and (\ref{condii}). 

As a criterion for succesful broadcasting we evaluate the fidelities of the output states with respect to the initial one $\kpsi $ of Eq. (\ref{ini}):
\beqa \label{fid}
F_{a_1b_1}&=&F(\rho_{a_1b_1},\proj{\psi}{\psi})\nonumber\\
&=&\frac{1}{4(1-3R+3R^2)^2}[4(1-2R)^2(1-R)^2+R^4+4(1-2R)(1-R)R^2(|\alpha |^4+|\beta |^4)];\nonumber\\
F_{cd}&=&F(\rho_{cd},\proj{\psi}{\psi})\nonumber\\
&=&\frac{1}{4(1-3R+3R^2)^2}[4R^2(1-R)^2+(1-2R)^4+4R(1-R)(1-2R)^2(|\alpha |^4+|\beta |^4)].
\eeqa
The fidelities of Eq. (\ref{fid}) are greater than 1/2, which represents the classical limit, i.e., the case when the initial entangled state is measured by the sender, then the outcome is sent to the receiver, who tries to reconstruct the state. The fidelities of the output states are plotted in Figs. 2 and 3 as a function of the coefficient $|\alpha |$ and the reflectivity of the beam splitter.

\begin{figure}
\includegraphics{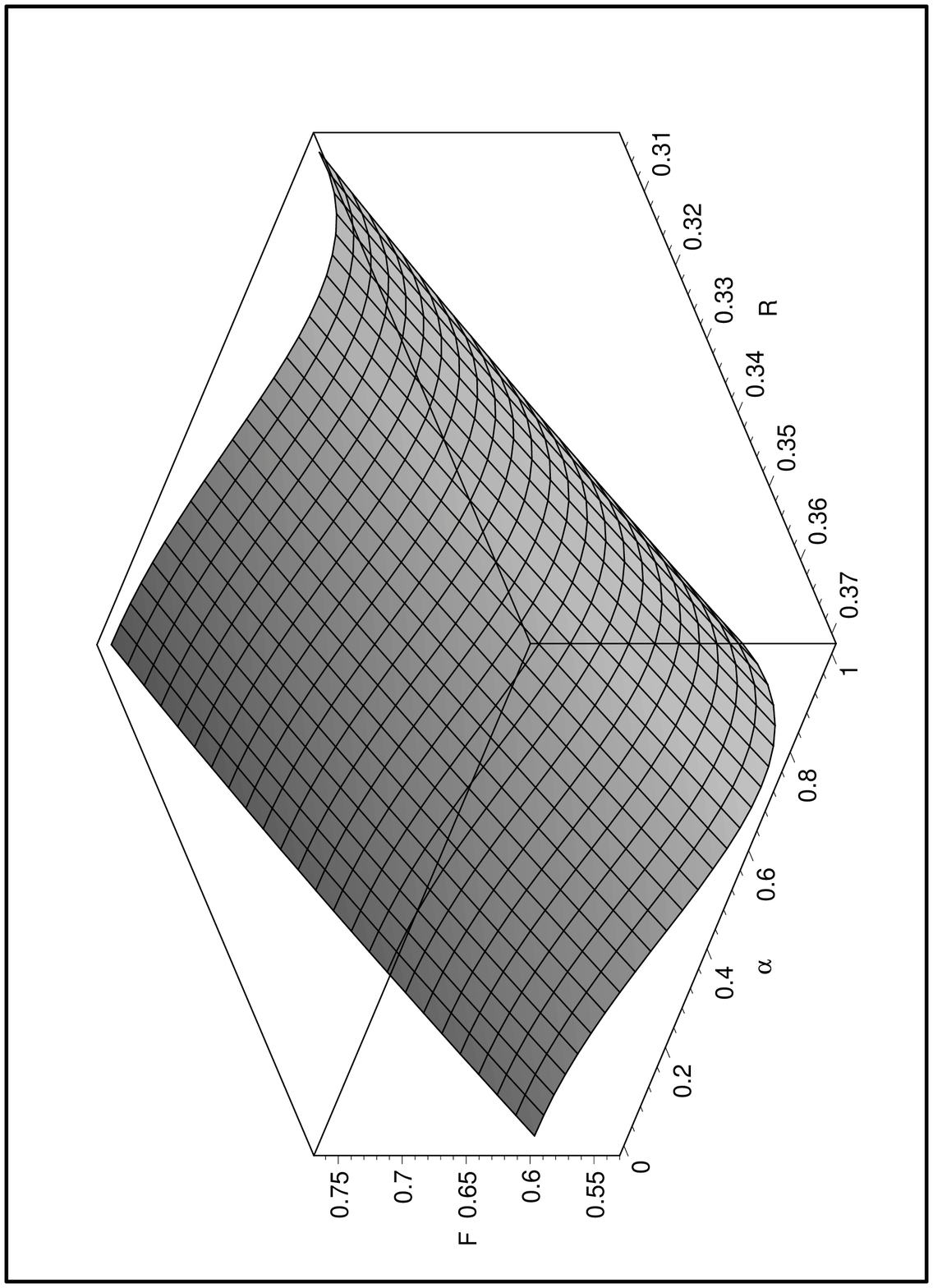}
\caption{The fidelity of the output state $\rho_{a_1b_1}$ with respect to the initial one as a function of the parameter $|\alpha |$, which characterizes the initial entangled state, and of the reflectivity $R$ of the beam splitters.}
\vspace{0.5cm}
\end{figure}  

\begin{figure}
\includegraphics{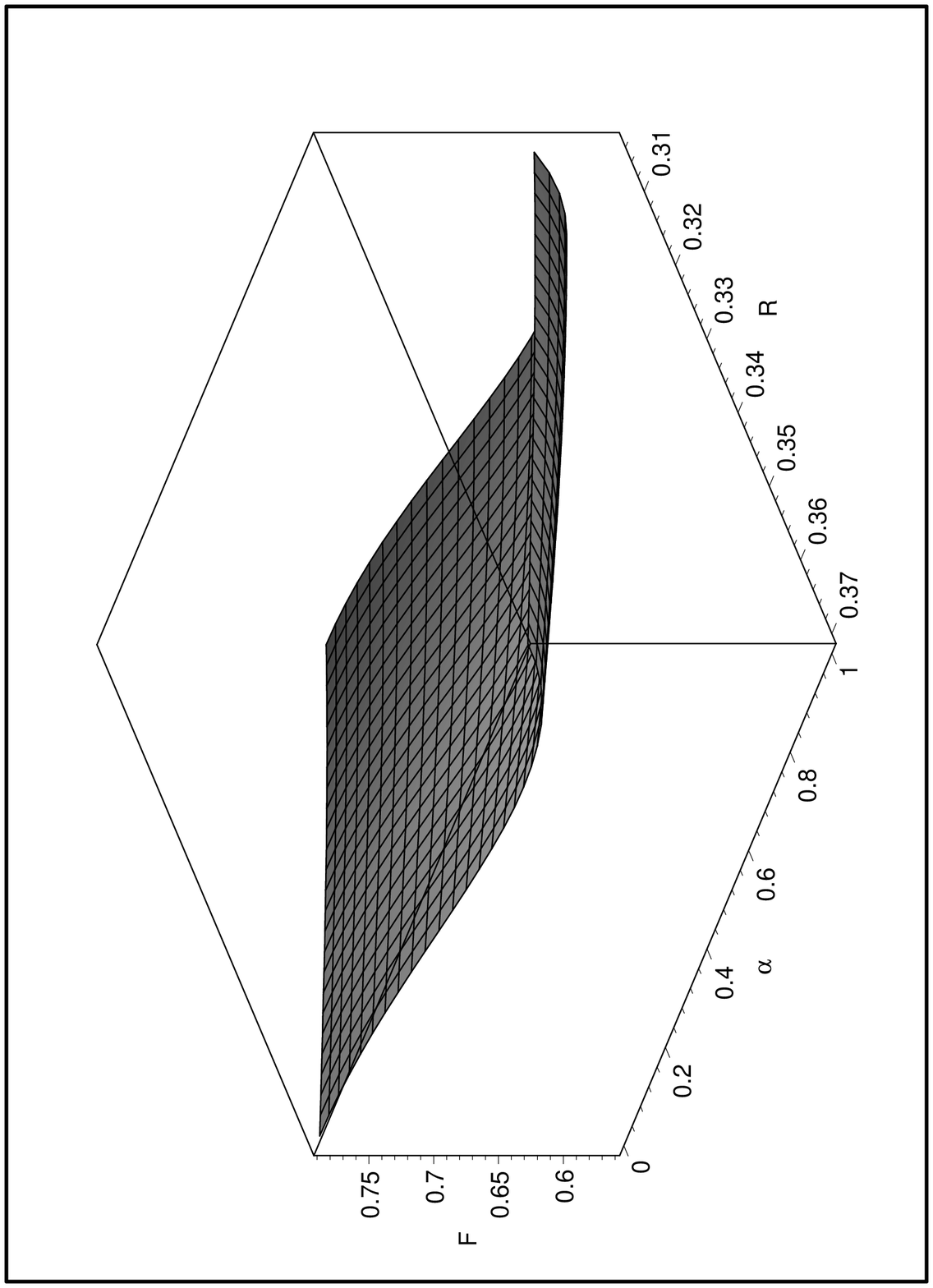}
\vspace{0.5cm}
\caption{The fidelity of the output state $\rho_{cd}$ with respect to the initial one as a function of the parameter $|\alpha |$, which characterizes the initial entangled state, and of the reflectivity $R$ of the beam splitters.}
\end{figure}  

The symmetric broadcasting of entanglement is obtained for $R=1/3$
\beqa \label{identice}
&&\rho_{{a_1}{b_1}}\left( \frac{1}{3}\right)=\rho_{cd}\left( \frac{1}{3}\right)=\frac{4}{9}\proj{\psi}{\psi}\nonumber\\
&&+\frac{8|\alpha |^2+1}{36}\proj{00}{00}+\frac{8|\beta |^2+1}{36}\proj{11}{11}\nonumber\\
&&+\frac{5}{36}(\proj{01}{01}+\proj{10}{10}).
\eeqa

We find from Eq. (\ref{actionBS}) that the entanglement is successful broadcast with the probability $\lambda_d$. For the symmetric protocol ($R=1/3$), the probability of unsuccessful outcomes, when the photons leave the beam splitters on the same arms, is
\beq
\lambda_s=\frac{5}{9}>\frac{4}{9}=\lambda_d.
\eeq
Althought our protocol for broadcasting of entanglement at the distance is a probabilistic one, this scheme presents the advantage that can be experimentally implemented since requires only linear optics for generating the final states.
Actually, in our protocol, Alice and Bob implement the partial optimal teleportation scheme on their particles of the initial entangled state $\ket{\psi}_{a_1b_1}$. If $\alpha$ and $\beta $ are real, the two final states (\ref{identice}) are identical with the ones obtained in the broadcasting of entanglement \cite{Buzek2} when each sender applies locally the symmetric optimal universal cloning machine $U(1/2)\otimes U(1/2)$ [see Eq. (\ref{Pauli})]. 
The fact that Alice and Bob share two singlet states with Charlie and Daniel enables to broadcast the entanglement at a distance.

Let us now discuss the possible experimental implementations of broadcasting of entanglement at the distance presented in Fig. 1. The two polarization maximally entangled states required in the scheme can be obtained by using type-II downconversion sources, such that the emerging photons have orthogonal polarizations \cite{Kwiat}, while for the initial state $\ket{\psi_{a_1b_1}}$, one can use type-I downconversion sources. The advantages of this method are that the sources are very easy to align and are very stable. On the other hand, one major disadvantage is that the polarization entanglement is subject to decoherence when it is transmitted over long-distance optical fibers.

By contrast, the time-bin encoding of entanglement distinguishes between the arrival times of the photons in the detectors, therefore this degree of freedom is suitable for quantum communication over long distances. Time-bin entangled photons at telecom wavelengths have been generated using a Ti:sapphire laser with femtosecond pulses by Marcikic {\it et al.} \cite{Gisin1}. A time-bin qubit is produced by passing the femtosecond pulse through an unbalanced interferometer with a relative phase $\varphi $ between the two arms:
\beq
\frac{1}{\sqrt 2}\left( \ket{1,0}-e^{i\varphi }\ket{0,1}\right),
\eeq
where $\ket{1,0}$ denotes the first time-bin when the photon passes through the short arm, while $\ket{0,1}$ is the second time-bin corresponding to the long arm. Time-bin entangled photons are generated by passing a time-bin qubit through a type-I nonlinear crystal, obtaining in this way a coherent superposition of photon pairs by downconversion,
\beq \label{timeentan}
\ket{\Lambda}=\frac{1}{\sqrt 2}\left( \ket{1,0}_A\ket{1,0}_B-e^{i\varphi }\ket{0,1}_A\ket{0,1}_B\right).
\eeq

Recently, the entanglement swapping protocol has been implemented by using two pairs of time-bin entangled photons emitted from separated sources at telecom wavelengths \cite{Gisin2}. This is the first entanglement swapping experiment performed over long distances (2 Km) in optical fibers.

Our experimental setup of Fig. 1 can be realized with the help of three time-bin entanglement sources. The state of Eq. (\ref{timeentan}) can be transformed to the singled state $\frac{1}{\sqrt 2}\left( \ket{1,0}_A\ket{0,1}_B-\ket{0,1}_A\ket{1,0}_B\right)$ by inserting switches and delays after the nonlinear crystal. Instead of using the beam splitter as is shown in the figure, we have to use a fiber coupler characterized by the variable coupling $R$. Therefore the broadcasting of entanglement over long distances can be implemented with the help of three separated sources of time-bin entangled photons at the telecom wavelengths.

%...................................

\subsection{Fidelity of teleportation of the output states obtained in the broadcasting of entanglement} 
Let us now analyze if the two output states of Eqs. (\ref{stfinala1}) and (\ref{stfinala2}) obtained in the process of broadcasting of entanglement violate a Bell inequality. An arbitrary two-level mixed state is written as \cite{Horodecki1}
\beq \label{romixt}
\rho =\frac{1}{4}\left( I\otimes I+\vec{r}\cdot \vec{\sigma}\otimes I+I\otimes \vec{s}\cdot \vec{\sigma}+\sum_{m,n=1}^3t_{mn}\sigma_m\otimes \sigma_n\right),
\eeq
where $\sigma_j$ are the Pauli matrices, $\vec{r}$ and $\vec{s}$ are real vectors, while $t_{mn}=\mbox{Tr}(\rho \sigma_m\otimes \sigma_n)$ form a real matrix called the correlation matrix, which is denoted by $T_\rho $. The Bell operator associated with the Clauser-Horne-Shimony-Holt (CHSH) inequality has the expression \cite{Horodecki1}
\beq
B=\vec{a}\cdot \vec{\sigma } \otimes (\vec{b}+\vec{b'})\cdot \vec{\sigma } +\vec{a'}\cdot \vec{\sigma } \otimes (\vec{b}-\vec{b'})\cdot \vec{\sigma },
\eeq
where $\vec{a}, \vec{a'}, \vec{b}, \vec{b'}$ are unit real vectors. The CHSH inequality is the following
\beq
|\expect{B}|\le 2.
\eeq
We define a symmetric matrix $U_\rho :=T_\rho ^TT_\rho $, whose eigenvalues are denoted by $\lambda_1, \lambda_2, \lambda_3$, with $\lambda_1\le \lambda_2\le \lambda_3$. Let us define $M(\rho )=\lambda_2+\lambda_3$. Then, Horodecki {\it et al.} have found the necessary and sufficient condition for violating the Bell-CHSH inequality by the mixed two-level states \cite{Horodecki1}.

{\it Theorem 1: The mixed state of Eq. (\ref{romixt}) violates the CHSH inequality if and only if $M(\rho)>1$}. 

For the final state $\rho_{a_1b_1}$, the correlation matrix $T$ reads
\beq
T_{\rho_{a_1b_1}}=\left( \begin{array}{ccc}
\frac{(1-2R)^2(1-R)^2}{(1-3R+3R^2)^2}(\alpha \beta^*+\alpha^*\beta )&i\frac{(1-2R)^2(1-R)^2}{(1-3R+3R^2)^2}(\alpha \beta^*-\alpha^*\beta )&0\\
i\frac{(1-2R)^2(1-R)^2}{(1-3R+3R^2)^2}(\alpha \beta^*-\alpha^*\beta )&-\frac{(1-2R)^2(1-R)^2}{(1-3R+3R^2)^2}(\alpha \beta^*+\alpha^*\beta )&0\\
0&0&\frac{(1-2R)^2(1-R)^2}{(1-3R+3R^2)^2}
\end{array}\right).
\eeq
The eigenvalues of $U_{\rho_{a_1b_1}}$ are 
\beqa
\lambda_1&=&\lambda_2=4\frac{(1-2R)^4(1-R)^4}{(1-3R+3R^2)^4}|\alpha |^2|\beta |^2,\nonumber\\
\lambda_3&=&\frac{(1-2R)^4(1-R)^4}{(1-3R+3R^2)^4},
\eeqa
therefore 
\beq \label{M}
M=\lambda_2+\lambda_3<1.
\eeq

For the state $\rho_{cd}$ we get
\beq
T_{\rho_{cd}}=\left( \begin{array}{ccc}
\frac{R^2(1-R)^2}{(1-3R+3R^2)^2}(\alpha \beta^*+\alpha^*\beta )&i\frac{R^2(1-R)^2}{(1-3R+3R^2)^2}(\alpha \beta^*-\alpha^*\beta )&0\\
i\frac{R^2(1-R)^2}{(1-3R+3R^2)^2}(\alpha \beta^*-\alpha^*\beta )&-\frac{R^2(1-R)^2}{(1-3R+3R^2)^2}(\alpha \beta^*+\alpha^*\beta )&0\\
0&0&\frac{R^2(1-R)^2}{(1-3R+3R^2)^2}
\end{array}\right).
\eeq
The eigenvalues of $U_{\rho_{cd}}$ are 
\beqa
\lambda_1&=&\lambda_2=4\frac{R^4(1-R)^4}{(1-3R+3R^2)^4}|\alpha |^2|\beta |^2,\nonumber\\
\lambda_3&=&\frac{R^4(1-R)^4}{(1-3R+3R^2)^4},
\eeqa
and they also satisfy Eq.(\ref{M}). Therefore by employing Theorem 1 we have found that the two final states $\rho_{a_1b_1}$ and $\rho_{cd}$ do not violate the CHSH inequality.

In the following we prove that although the two final states do not violate the Bell inequality, they are useful in quantum information processes, namely as quantum channels in the standard teleportation protocol.

 Assume that we want to teleport an arbitrary two-level state given by $P_\phi=\half (I+\vec{a}\cdot \vec{\sigma})$. In the original teleportation scheme, the information is sent over a quantum channel formed by the singlet state, and finally the state is faithfully restored (with unit fidelity) \cite{Bennett}. If the sender is allowed to measure in the Bell basis, while the receiver applies any unitary operations, then the scheme is called standard teleportation \cite{Horodecki2,Horodecki3}. Horodecki {\it et al.} have investigated the standard teleportation which uses mixed inseparable states as quantum channels. They have introduced as a measure of efficiency of transmission, the fidelity of teleportation, which is given by
\beq
{\cal F}=\int \sum_kp_k\mbox{Tr}(\rho_kP_\phi )dM(\phi ),
\eeq
where $\rho_k$ is the output state obtained with the probability $p_k$. 

Horodecki {\it et al.} have defined a quantity associated with the mixed state of Eq. (\ref{romixt}): $N(\rho ):=\mbox{Tr}\sqrt{T_\rho^TT_\rho }$ and they have proved the following \cite{Horodecki2}:

{\it Theorem 2: Any two-level mixed state is useful for the standard teleportation if and only if $N(\rho )>1$, and the fidelity of teleportation is given by }
\beq
{\cal F}_{max}=\half \left[ 1+\frac{1}{3}N(\rho )\right].
\eeq

Now suppose we want to use the output mixed states of Eqs. (\ref{stfinala1}) and (\ref{stfinala2}) obtained in the broadcasting of entanglement as quantum channels for the standard teleportation. We find
\beqa \label{nfinale}
N(\rho_{a_1b_1})&=&\frac{(1-2R)^2(1-R)^2}{(1-3R+3R^2)^2}(4|\alpha ||\beta |+1),\nonumber\\
N(\rho_{cd})&=&\frac{R^2(1-R)^2}{(1-3R+3R^2)^2}(4|\alpha ||\beta |+1).
\eeqa

Let us investigate the case when the initial state of Eq. (\ref{ini}) to be broadcast is a maximally entangled one, i.e., $\alpha =\frac{1}{\sqrt 2}$. According to Eq. (\ref{nfinale}), we obtain for the output states $\rho_{a_1b_1}$ and $\rho_{cd}$:
\beqa 
N(\rho_{a_1b_1})&=&3\frac{(1-2R)^2(1-R)^2}{(1-3R+3R^2)^2},\nonumber\\
N(\rho_{cd})&=&3\frac{R^2(1-R)^2}{(1-3R+3R^2)^2},
\eeqa
which are greater than 1 for $R\in \left( \half -\frac{1}{6}\sqrt{-9+6\sqrt 3},x\right)$ as it is imposed by Eqs. (\ref{condi}) and (\ref{condii}). This means that the two states are useful for the standard teleportation according to the second theorem.

We analyze now another interesting case, namely when the states are symmetrically broadcast: $R=\frac{1}{3}$:
\beq
N(\rho_{a_1b_1})=N(\rho_{cd})=\frac{4}{9}(4|\alpha ||\beta |+1)>1
\eeq
if $|\alpha |\in \left[ \sqrt{\half -\frac{1}{16}\sqrt{39}},\sqrt{\half +\frac{1}{16}\sqrt{39}}\right]$ [see Eq. (\ref{condi}) for $R=1/3$]. The fidelity of teleportation is greater than the one obtained using a separable bipartite state as quantum channel,
\beq
F=\half \left[ 1+\frac{4}{27}\left( 4|\alpha ||\beta |+1\right) \right]>F_{classic}=\frac{2}{3}.
\eeq
Despite the fact that the Bell inequalities are not violated, the output states generated in the broadcasting of entanglement can be used as quantum channels for the standard teleportation. 
A similar result was found by Popescu for the Werner state \cite{Popescu,Werner2},
\beq
\rho_W=\frac{1}{8}I+\half \proj{\psi^-}{\psi^-},
\eeq
where one shows that the inseparability required in quantum teleportation is not equivalent to the Bell's inequality violation.

%....................................
\section{Telecloning of entanglement}

Having investigated the broadcasting of one entangled state to
two locations,  let us now investigate the following scenario:
assume that two spatially separated observers, $A_1$ and $A_2$,
hold an entangled state and they wish to send two copies of this
state to two pairs of observers also located at different places
$B_1$, $B_2$, and
 $B_3$, $B_4$, respectively.

 Let the initial entangled state be (note again that $\alpha$ and
 $\beta$ are unknown)
\beq \label{initial}
 \kpsi_{A_1A_2}=\alpha \ket{00}+\beta \ket{11}.
\eeq

 A simple way for $A_1$ and $A_2$ to perform this scenario is as follows: $A_2$
 teleports his particle of the initial entangled state to $A_1$ using the standard
 teleportation protocol with the help of an additional maximally entangled state \cite{Bennett}.
 Further $A_1$ generates two nonlocal asymmetric clones as was shown in Ref. \cite{Ghiu},
 and finally he teleports the four particles to the receivers.
 The entanglement required is $E_1=log_2 2=1$
 between $A_1$ and $A_2$, and $E_2=4log_2 2=4$ between $A_1$ and $B_1, B_2, B_3, B_4$.
 Also 2-bit clssical communication from $A_2$ to $A_1$, and 8-bit classical communication from
 $A_1$ to  $B_1, B_2, B_3, B_4$ is used in this scenario.

Here we propose a different scheme, namely telecloning of entanglement for performing this protocol, which simultaneously copy and transfer the information encoded in an arbitrary entangled state at a distance.

In order to define the quantum channel we write the action of the
optimal universal asymmetric cloning machine (\ref{general})
   characterized by $d=4$ on the state $\ket{\psi}\ket{00}\ket{00}$:

\beq  U\kpsi_{12} \ket{00}_{34}\ket{00}_{56}=\alpha
\ket{\eta_0}+\beta \ket{\eta_1}, \eeq where \beqa \label{eta0}
&&\ket{\eta_0}:=\frac{1}{\sqrt{1+3(p^2+q^2)}}(\ket{000000}+p\ket{000101}\nonumber\\
&&+p\ket{001010}+p\ket{001111}+q\ket{010001}\nonumber\\
&&+q\ket{100010}+q\ket{110011}), \eeqa and \beqa \label{eta1}
&&\ket{\eta_1}:=\frac{1}{\sqrt{1+3(p^2+q^2)}}(\ket{111111}+p\ket{110000}\nonumber\\
&&+p\ket{110101}+p\ket{111010}+q\ket{001100}\nonumber\\
&&+q\ket{011101}+q\ket{101110}). \eeqa In the above expressions we
have replaced $\ket{0}$ by $\ket{00}$, $\ket{1}$ by $\ket{01}$,
$\ket{2}$ by $\ket{10}$,
  $\ket{3}$ by $\ket{11}$.

Next we apply the many-to-many teleportation protocol proposed by
one of us \cite{Ghiu} in order to encode the information of the
initial state $\ket{\psi}_{A_1A_2}$ to a six-particle state shared
by six observers spatially separated, 
\beq \ket{\Pi}=\alpha
\ket{\eta_0}_{B_1B_2B_3B_4B_5B_6}+\beta
\ket{\eta_1}_{B_1B_2B_3B_4B_5B_6}, \eeq where the states
$\ket{\eta_0}$ and $\ket{\eta_1}$ are defined in Eqs.
(\ref{eta0}) and (\ref{eta1}).

The quantum channel is a maximally entangled state shared between the senders and receivers:
\beqa \label{telecstate}
&&\ket{\xi}=\frac{1}{\sqrt 2}\ket{00}_{A'_1A'_2}\ket{\eta_0}_{B_1B_2B_3B_4B_5B_6}\nonumber\\
&&+\frac{1}{\sqrt 2}\ket{11}_{A'_1A'_2}\ket{\eta_1}_{B_1B_2B_3B_4B_5B_6},
\eeqa
where the particles denoted by $A'_1, A'_2$ belong to the senders
$A_1, A_2$, while $B_5, B_6$ are the two observers who hold the ancillas.

The state of the whole system is
\beqa
&&\ket{\psi}\ket{\xi}=\frac{1}{2\sqrt 2}[\ket{\Phi^+}\ket{\Phi^+}(\alpha \ket{\eta_0}+\beta \ket{\eta_1})\nonumber\\
&&+\ket{\Phi^+}\ket{\Phi^-}(\alpha \ket{\eta_0}-\beta \ket{\eta_1})\nonumber\\
&&+\ket{\Phi^-}\ket{\Phi^+}(\alpha \ket{\eta_0}-\beta \ket{\eta_1})+\ket{\Phi^-}\ket{\Phi^-}(\alpha \ket{\eta_0}+\beta \ket{\eta_1})\nonumber\\
&&+\ket{\Psi^+}\ket{\Psi^+}(\alpha \ket{\eta_1}+\beta \ket{\eta_0})+\ket{\Psi^+}\ket{\Psi^-}(\alpha \ket{\eta_1}-\beta \ket{\eta_0})\nonumber\\
&&+\ket{\Psi^-}\ket{\Psi^+}(\alpha \ket{\eta_1}-\beta \ket{\eta_0})+\ket{\Psi-}\ket{\Psi^-}(\alpha \ket{\eta_1}+\beta \ket{\eta_0})].
\eeqa

\begin{figure}
\includegraphics{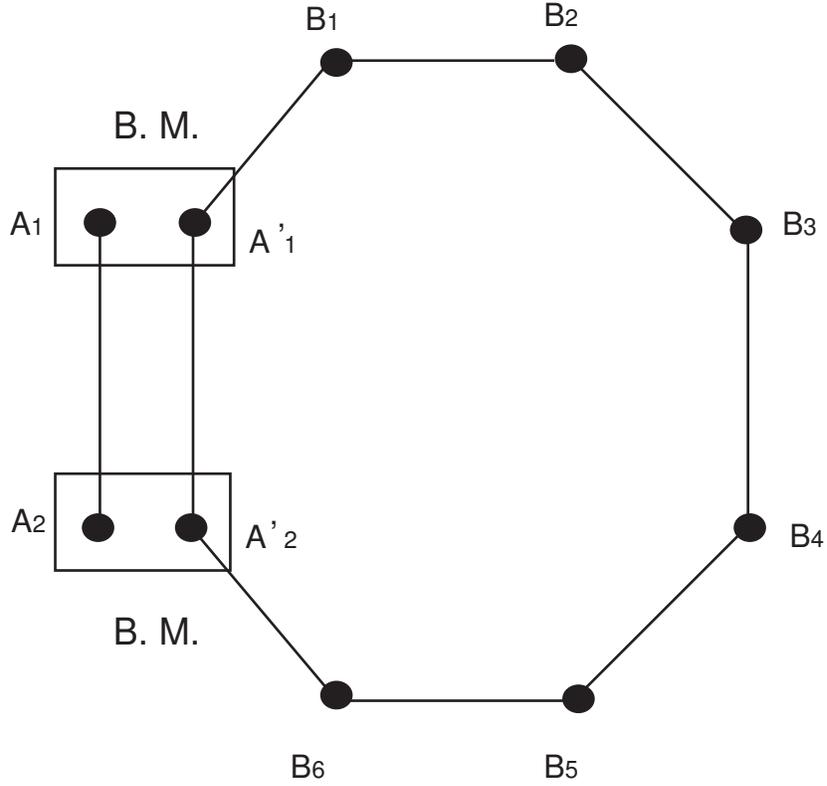}
 \caption{Schematic of the asymmetric telecloning of
entanglement. This protocol is obtained by combining the
many-to-many teleportation and the asymmetric optimal cloning. The
quantum channel used is a maximally entangled eight-particle
state. Each sender $A_1, A_2$ has to perform locally a Bell-type
measurement (B.M.) and announce the outcome to the receivers, who
perform a local recovery unitary operation obtaining the
information of the initial unknown state. The final inseparable
states states are shared by the receivers $B_1-B_2$ and $B_3-B_4$,
respectively. The lines represent the entanglement.}
\end{figure}

The protocol consists of three steps:\\
(i) Each sender performs a measurement of his particles in the Bell basis
 as is shown in Fig. 4.\\
(ii) The senders communicate the outcomes to the receivers.\\
(iii) The receivers apply local unitary operations depending on the outcomes of the senders' measurements. Let us analyze the case in which the Bell-state analyzers give the result $\ket{\Phi^+}\ket{\Phi^-}$, which occurs with a probability 1/8. Therefore the state of the receivers is projected on $\alpha \ket{\eta_0}-\beta \ket{\eta_1}$. Then the receivers $B_1, B_3, B_5$ perform the unitary operation $\sigma_z$, while the receivers $B_2, B_4, B_6$ do nothing, in order to retrieve the information contained in the initial state.
In Table I we have shown the local recovery unitary operations that have to be performed for each outcome of the Bell measurements.

\begin{table}
\caption{The local recovery unitary operations that have to be applied by the receivers, which depend on the outcomes of the senders' measurements.}
\begin{tabular}[t]{|c|c|}
\hline
Outcome&Local recovery unitary operation \\ \hline
\hline
$\ket{\Phi^+}\ket{\Phi^+}$&$I\otimes I\otimes I\otimes I\otimes I\otimes I$\\  \hline
$\ket{\Phi^+}\ket{\Phi^-}$&$\sigma_z\otimes I\otimes \sigma_z\otimes I \otimes \sigma_z\otimes I$\\ \hline
$\ket{\Phi^-}\ket{\Phi^+}$&$\sigma_z\otimes I\otimes \sigma_z\otimes I \otimes \sigma_z\otimes I$\\ \hline
$\ket{\Phi^-}\ket{\Phi^-}$&$I\otimes I\otimes I\otimes I\otimes I\otimes I$\\ \hline
$\ket{\Psi^+}\ket{\Psi^+}$&$\sigma_x\otimes \sigma_x\otimes \sigma_x\otimes \sigma_x\otimes \sigma_x\otimes \sigma_x$\\ \hline
$\ket{\Psi^+}\ket{\Psi^-}$&$\sigma_x \sigma_z\otimes \sigma_x\otimes \sigma_x \sigma_z\otimes \sigma_x\otimes\sigma_x \sigma_z\otimes \sigma_x$\\ \hline
$\ket{\Psi^-}\ket{\Psi^+}$&$\sigma_x \sigma_z\otimes \sigma_x\otimes \sigma_x \sigma_z\otimes \sigma_x\otimes\sigma_x \sigma_z\otimes \sigma_x$\\ \hline
$\ket{\Psi^-}\ket{\Psi^-}$&$\sigma_x\otimes \sigma_x\otimes \sigma_x\otimes \sigma_x\otimes \sigma_x\otimes \sigma_x$\\ \hline
\end{tabular}
\end{table}

Therefore we have faithfully encoded the information of the initial two-particle state (\ref{initial}) into the six-particle state:
\beq
\ket{\Pi}=\alpha \ket{\eta_0}_{B_1B_2B_3B_4B_5B_6}+\beta \ket{\eta_1}_{B_1B_2B_3B_4B_5B_6}.
\eeq

We obtain two entangled pairs shared by two pairs of receivers,
$B_1-B_2$ and $B_3-B_4$, characterized by the following density
operators: \beq \rho_{B_1B_2}=\frac{1}{1+3(p^2+q^2)}\left[ \left(
1-q^2+3p^2\right) \proj{\psi}{\psi}+q^2I\right], \eeq \beq
\rho_{B_3B_4}=\frac{1}{1+3(p^2+q^2)}\left[ \left(
1-p^2+3q^2\right) \proj{\psi}{\psi}+p^2I\right], \eeq which are
the final states generated by nonlocal optimal universal
asymmetric cloning machines.

In the telecloning of entanglement one needs the entanglement $E=1$ between the senders and receivers, and only 4 classical bits to be transmitted from $A_1$ and $A_2$ to the receivers. Therefore this protocol is more efficient than the first one in terms of entanglement resources and the amount of classical bits.

It should be emphasized that the method which we have introduced above for telecloning of entanglement requires an eight-particle entangled state as a resource. If our scheme would be implemented using an eight-photon entangled state, then the efficiency decreases up to 50\% due to impossibility of performing perfect Bell state measurements using linear optics. Therefore it could be more interesting that the protocol to be implemented with the help of atomic entangled states or nuclear spin entangled states.

\section{Discussions and Conclusions}

In conclusion, we have generalized partial teleportation and
telecloning of quantum bits to the partial cloning of entangled
states (more generally quantum registers). Given the resources of
entangled state, our first scheme for splitting of entanglement
requires only linear optical components such as beamsplitters and
single-photon detectors. In particular for $R=1/3$, our scheme is an implementation of the protocol prezented by Buzek {\it et al.} \cite{Buzek2}. By using two theorems introduced by Horodecki {\it et al.} based on the correlation matrix, we prove that the output states do not violate the CHSH-Bell inequality, but they are useful for the standard teleportation.

Then, we have introduced a protocol
called telecloning of entanglement, which distributes the two
clones of an entangled state to two pairs of observers spatially
separated. We use as channel a maximally entangled
state shared between the senders and receivers. In the telecloning of entanglement one consumes only one e-bit of entanglement between the senders and receivers, and 4 classical bits, being more efficient than the straightforward protocol where one of the senders locally generates the clones, and then teleports them to the receivers, scheme which requires one e-bit of entanglement between the senders and 4 e-bit of entanglement between this sender and receivers, and in addition 10 classical bits.

The results on the entanglement transfer of
this paper may be of relevance for quantum communication (i.e.,
teleportation, quantum repeaters) as well as for distributed
quantum computing \cite{Hardy}.

\section*{Acknowledgments}
I. Ghiu thanks Professor Tudor A. Marian from the University of
Bucharest for fruitful discussions. This work was supported by
the Romanian CNCSIS through a
grant for the University of Bucharest, 
 the
Swedish foundation for strategic research-SSF, and the Royal Swedish
Academy of Sciences - KVA.

\appendix
%\renewcomand\theequation{\appendix \arabic{equation}}

\section{The inseparability of the final states of the broadcasting of entanglement}

Let us now analyze the inseparability of the final states
described in Sec. II B. The normalized final state after applying the
operator $\Pi_{a_1a_2}\otimes I_c\otimes \Pi_{b_1b_2}\otimes I_d$
on the input state (\ref{init}) is
 \beqa \label{fi}
&&\ket{\phi_d}=\frac{1}{\sqrt{\lambda_d}}\{\alpha [(1-2R)^2\ket{00}_{{a_1}{a_2}}\ket{00}_{{b_1}{b_2}}\ket{11}_{cd}\nonumber\\
&&+(1-R)^2\ket{01}_{{a_1}{a_2}}\ket{01}_{{b_1}{b_2}}\ket{00}_{cd}\nonumber\\
&&+R^2\ket{10}_{{a_1}{a_2}}\ket{10}_{{b_1}{b_2}}\ket{00}_{cd}\nonumber\\
&&-(1-2R)(1-R)\ket{00}_{{a_1}{a_2}}\ket{01}_{{b_1}{b_2}}\ket{10}_{cd}\nonumber\\
&&-(1-2R)(1-R)\ket{01}_{{a_1}{a_2}}\ket{00}_{{b_1}{b_2}}\ket{01}_{cd}\nonumber\\
&&+R(1-2R)\ket{00}_{{a_1}{a_2}}\ket{10}_{{b_1}{b_2}}\ket{10}_{cd}\nonumber\\
&&+R(1-2R)\ket{10}_{{a_1}{a_2}}\ket{00}_{{b_1}{b_2}}\ket{01}_{cd}\nonumber\\
&&-R(1-R)\ket{01}_{{a_1}{a_2}}\ket{10}_{{b_1}{b_2}}\ket{00}_{cd}\nonumber\\
&&-R(1-R)\ket{10}_{{a_1}{a_2}}\ket{01}_{{b_1}{b_2}}\ket{00}_{cd}]\nonumber\\
&&+\beta [(1-2R)^2\ket{11}_{{a_1}{a_2}}\ket{11}_{{b_1}{b_2}}\ket{00}_{cd}\nonumber\\
&&+(1-R)^2\ket{10}_{{a_1}{a_2}}\ket{10}_{{b_1}{b_2}}\ket{11}_{cd}\nonumber\\
&&+R^2\ket{01}_{{a_1}{a_2}}\ket{01}_{{b_1}{b_2}}\ket{11}_{cd}\nonumber\\
&&-(1-2R)(1-R)\ket{11}_{{a_1}{a_2}}\ket{10}_{{b_1}{b_2}}\ket{01}_{cd}\nonumber\\
&&-(1-2R)(1-R)\ket{10}_{{a_1}{a_2}}\ket{11}_{{b_1}{b_2}}\ket{10}_{cd}\nonumber\\
&&+R(1-2R)\ket{11}_{{a_1}{a_2}}\ket{01}_{{b_1}{b_2}}\ket{01}_{cd}\nonumber\\
&&+R(1-2R)\ket{01}_{{a_1}{a_2}}\ket{11}_{{b_1}{b_2}}\ket{10}_{cd}\nonumber\\
&&-R(1-R)\ket{10}_{{a_1}{a_2}}\ket{01}_{{b_1}{b_2}}\ket{11}_{cd}\nonumber\\
&&-R(1-R)\ket{01}_{{a_1}{a_2}}\ket{10}_{{b_1}{b_2}}\ket{11}_{cd}]\},
\eeqa
where $\lambda_d=4(1-3R+3R^2)^2$.

Applying the Peres-Horodecki theorem \cite{Peres,Horodecki} we find that the state $\rho_{{a_1}{b_1}}$ of Eq. (\ref{stfinala1}) is inseparable if
\beq \label{r1}
R\in [0,x),
\eeq
 where $x=0.3608506129...$ is one solution of the equation $3R^4-18R^3+24R^2-12R+2=0$. Also we find for which values of $\alpha $ this state is inseparable
\beqa \label{alfa1}
&&\frac{1}{2}\left( 1-\sqrt{1-\frac{R^4(2-6R+5R^2)^2}{4(1-2R)^4(1-R)^4}}\right) \le |\alpha |^2\nonumber\\
&&\le \frac{1}{2}\left( 1+\sqrt{1-\frac{R^4(2-6R+5R^2)^2}{4(1-2R)^4(1-R)^4}}\right).
\eeqa

The state $\rho_{cd}$ of Eq. (\ref{stfinala2}) is inseparable if
\beq \label{r2}
R\in \left( \half -\frac{1}{6}\sqrt{-9+6\sqrt 3},\half +\frac{1}{6}\sqrt{-9+6\sqrt 3}\right),
\eeq
and for $\alpha $ which satisfies
\beqa \label{alfa2}
&&\frac{1}{2}\left( 1-\sqrt{1-\frac{(1-2R)^4(1-2R+2R^2)^2}{4R^4(1-R)^4}}\right)\le |\alpha |^2\nonumber\\
&&\le \frac{1}{2}\left( 1+\sqrt{1-\frac{(1-2R)^4(1-2R+2R^2)^2}{4R^4(1-R)^4}}\right).
\eeqa

We have to find the conditions when the states $\rho_{a_1c}, \rho_{b_1d}$ are separable:
\beqa
&&\rho_{a_1c}= \rho_{b_1d}=\frac{1}{2(1-3R+3R^2)}\{(1-R)^2|\alpha |^2\proj{00}{00}\nonumber\\
&&+(1-R)^2|\beta |^2\proj{11}{11}+2R(1-2R)\proj{\psi^+}{\psi^+}\nonumber\\
&&+[(1-4R+3R^2)|\alpha |^2+R(3R-1)]\proj{01}{01}\nonumber\\
&&+[(1-4R+3R^2)|\beta |^2+R(3R-1)]\proj{10}{10}\},
\eeqa
where $\ket{\psi^+}=1/\sqrt 2(\ket{01}+\ket{10})$ is a component of the Bell basis. These conditions are
\beq \label{r3}
R\in \left[ 0,\frac{1}{\sqrt 3}\right]
\eeq
and
\beqa \label{alfa3}
&&\half \left( 1-\sqrt{1-\frac{4R^2(1-2R)^2}{(1-R)^4}}\right)\le |\alpha |^2\nonumber\\
&&\le \half \left( 1+\sqrt{1-\frac{4R^2(1-2R)^2}{(1-R)^4}}\right).
\eeqa

In conclusion, combining Eqs. (\ref{r1}), (\ref{r2}), (\ref{r3}), and (\ref{alfa1}), (\ref{alfa2}), (\ref{alfa3}) we obtain that

(i) if
$R\in \left( \half -\frac{1}{6}\sqrt{-9+6\sqrt 3},\frac{1}{3}\right] $,
\beqa \label{condi}
&&\frac{1}{2}\left( 1-\sqrt{1-\frac{(1-2R)^4(1-2R+2R^2)^2}{4R^4(1-R)^4}}\right)\le |\alpha |^2\nonumber\\
&&\le \frac{1}{2}\left( 1+\sqrt{1-\frac{(1-2R)^4(1-2R+2R^2)^2}{4R^4(1-R)^4}}\right)
\eeqa

(ii) if $R\in \left( \frac{1}{3},x\right)$,
\beqa \label{condii}
&&\frac{1}{2}\left( 1-\sqrt{1-\frac{R^4(2-6R+5R^2)^2}{4(1-2R)^4(1-R)^4}}\right) \le |\alpha |^2\nonumber\\
&&\le \frac{1}{2}\left( 1+\sqrt{1-\frac{R^4(2-6R+5R^2)^2}{4(1-2R)^4(1-R)^4}}\right),
\eeqa
then the states $\rho_{{a_1}c}$, $\rho_{{b_1}d}$ are separable when the states $\rho_{{a_1}{b_1}}$, $\rho_{cd}$ are inseparable.

\end{document}